**Semiclassical rates for tunnel ionization from power-law potentials induced by a constant or low-frequency electric field**

**A.M. Ishkhanyan[1,2] and V.P. Krainov[3]**

[1]Institute for Physical Research of NAS RA, Ashtarak, 0203, Armenia
[2]Physic and Technology Institute, National Research University Tomsk Polytechnic University, Tomsk, 634050 Russia
[3]Moscow Institute for Physics and Technology, Dolgoprudny, 141700, Russia

The probabilities of tunnel ionization of particles confined by one-dimensional power-law and logarithmic potentials are calculated for constant and low-frequency electric fields.

## 1. Introduction

The probability of tunnel ionization of quantum systems induced by a constant external field is described with an exponential accuracy by the barrier penetrability [1,2]. For example, if an electron is in a bound state of an atomic system with the energy $E_n$, then for the applied weak electric field with intensity $F$ the semiclassical WKB penetrability has a rather simple form (we put $m=\hbar=e=1$):

$$D \sim \exp\left(-\frac{2\left(2\left|E_n\right|\right)^{3/2}}{3F}\right) \ll 1 . \tag{1}$$

However, the pre-exponential factor in this relation may be both very large and very small. Also, the form of this factor depends on the potential where the electron (or other particle) is confined. For example, the ionization probability per unit time for ionization from the ground state of the hydrogen atom is given as [1]

$$w=\frac{4}{F}\exp\left(-\frac{2}{3F}\right), \quad F \ll 1 . \tag{2}$$

And for ionization from the ground state with energy $-\kappa^2/2$ in the spherical short-range well, the probability of ionization per unit time is [1]

$$w=\frac{F}{2\kappa}\exp\left(-\frac{2\kappa^3}{3F}\right), \quad F \ll \kappa^3 . \tag{3}$$

In the present paper we discuss the tunnel ionization of semiclassical bound states $\{E_n, \phi_n(x)\}$ of quantum systems confined by the one-dimensional power-law potential

$$V(x) = -\frac{V_0}{x^s}, \quad x > 0, \quad 0 < s < 2. \tag{4}$$

First, we turn to a simpler problem, when the external electric field does not depend on time and is defined by the constant field intensity $F$ (figure 1):

$$V = -\frac{V_0}{x^s} - Fx. \tag{5}$$

The transition to the variable but low-frequency field could be done from the derived expressions for the probability rate by a mere averaging over the period of the electromagnetic field. We use the units $m = \hbar = e = V_0 = 1$.

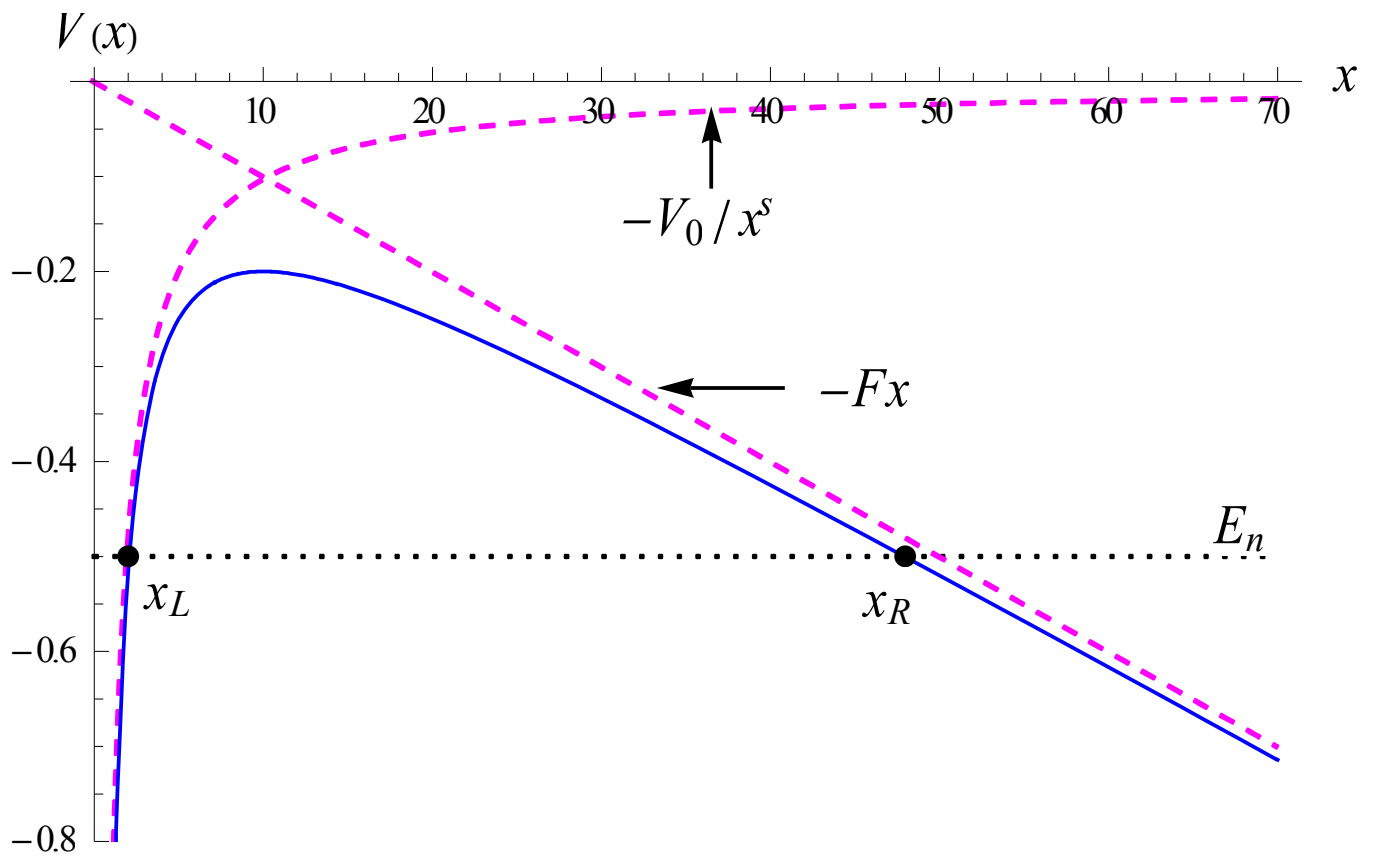

Fig. 1. The potential (5) with $s = 1$, $F = 0.01$. The left-hand and right-hand turning points are marked as $x_L$ and $x_R$.

**2. Case** $1 < s < 2$

The stationary Schrödinger equation is written as

$$\left(\frac{1}{2}\hat{p}^2 - \frac{1}{x^s} - Fx - E\right)\phi(x) = 0. \tag{6}$$

We search for the solution of this equation obeying the following conditions for $F \to 0$:

$$E \to E_n, \quad \phi(x) \to \phi_n(x). \tag{7}$$

We assume that the field is weak in comparison with the atomic field. In the classically accessible region the wave-functions $\phi_n(x)$ are presented in [3,4]. The unperturbed WKB wave-function in the initial region under the potential barrier (beyond the classical turning point $x_0 = |E_n|^{-1/s}$), where the electric field can be neglected, has the form

$$\phi_n(x) \to \frac{A}{2\sqrt{|p_0(x)|}} \exp\left( -\int_{x_0}^{x} |p_0(x')| dx' \right), \quad x > x_0, \tag{8}$$

$$|p_0(x)| = \sqrt{2\left( |E_n| - \frac{1}{x^s} \right)}, \quad |E_n| = \frac{1}{x_0^s}. \tag{9}$$

The energies $E_n$ are known [3,4]. The normalization factor $A$ in equation (8) is defined by the classically accessible region:

$$\phi_n(x) = \frac{A}{\sqrt{p_0(x)}} \cos\left( \int_{x}^{x_0} p_0(x') dx' - \frac{\pi}{4} \right), \quad A^2 \int_0^{x_0} \frac{dx}{2 p_0(x)} = 1. \tag{10}$$

Calculating the integral, we get

$$A^2 = 2 x_0^{-1-s/2} \sqrt{\frac{2}{\pi}} \Gamma\left( \frac{1}{s} \right) / \Gamma\left( \frac{1}{2} + \frac{1}{s} \right). \tag{11}$$

Consider the under-barrier region $x \sim |E_n| / F$. Neglecting the quadratic Stark shift, we replace $E \to E_n$. The momentum when the particle moves along the *x*-axes under the barrier is

$$|p(x)| = \sqrt{2\left( |E_n| - \frac{1}{x^s} - Fx \right)}. \tag{12}$$

And the quasiclassical wave-function is the analytic continuation of the quasiclassical function (10) into the region, where the electric field becomes significant:

$$\phi(x) \to \frac{A}{2\sqrt{|p(x)|}} \exp\left( -\int_{x_0}^{x} |p(x')| dx' \right), \quad x_0 < x < x_1 = \frac{|E_n|}{F}. \tag{13}$$

If the electric field is weak, the right-hand turning point $x_1$ is determined in the zero-order approximation by neglecting the atomic potential:

$$\frac{1}{x_1^s} \ll F x_1, \quad F\left( \frac{|E_n|}{F} \right)^{s+1} \gg 1, \quad F \ll |E_n|^{1+1/s} \tag{14}$$

(below we take into account the shift of the turning points wherever it is significant). The ionization rate is given by the density of the probability current at the exit from under the barrier. According to equation (13), it is equal to

$$w=\frac{A^2}{4}\exp\left(-2\int_{x_0}^{x_1}|p(x)|\,dx\right). \qquad (15)$$

Since the exponent index is much larger than unity, not only the leading term but also the next one should be retained in the integral (15) to obtain the correct pre-exponential factor in this expression for the probability. We will see that the leading term is inversely proportional to the field intensity, while the next term does not depend on that. This justifies the neglecting the quadratic Stark shift.

Neglecting the atomic potential, we get the following leading term in the exponential involved in equation (15):

$$I_{s1}=-\frac{2\left(2|E_n|\right)^{3/2}}{3F}. \qquad (16)$$

This result is what was expected as this index of the exponential function is the one generally encountered when discussing the tunneling from the common atomic potentials.

The next term gives

$$I_{s2}=2\int_0^{x_0}\sqrt{2\left(|E_n|-Fx\right)}dx\approx 2\int_0^{x_0}\sqrt{2\left(|E_n|\right)}dx=\frac{2\sqrt{2}}{|E_n|^{1/s-1/2}}. \qquad (17)$$

Finally, the last term is (here we put $F=0$):

$$I_{s3}=-2\int_{x_0}^{x_1}\left(\sqrt{2\left(|E_n|-\frac{1}{x^s}-Fx\right)}-\sqrt{2\left(|E_n|-Fx\right)}\right)dx=$$
$$=4\int_{x_0}^{x_1}\frac{dx}{\left(\sqrt{2\left(|E_n|-\frac{1}{x^s}-Fx\right)}+\sqrt{2\left(|E_n|-Fx\right)}\right)x^s}\approx\int_{x_0}^{\infty}\frac{2\sqrt{2}dx}{\left(\sqrt{|E_n|-\frac{1}{x^s}}+\sqrt{|E_n|}\right)x^s}. \qquad (18)$$

The change of the variable $x=y/|E_n|^{1/s}$ leads to a dimensionless integral, which converges for $s>1$:

$$I_{s3}=\frac{2\sqrt{2}}{|E_n|^{1/s-1/2}}f(s),\quad f(s)=\int_1^{\infty}\frac{dy}{\left(\sqrt{1-1/y^s}+1\right)y^s}. \qquad (19)$$

So, the probability of ionization rate is given by the expression

$$w=|E_n|^{1/2+1/s}\sqrt{\frac{1}{2\pi}}\frac{\Gamma\left(\frac{1}{s}\right)}{\Gamma\left(\frac{1}{2}+\frac{1}{s}\right)}\exp\left(-\frac{2\left(2|E_n|\right)^{3/2}}{3F}+\frac{2\sqrt{2}\left(f(s)+1\right)}{|E_n|^{1/s-1/2}}\right). \qquad (20)$$

The function $f(s)$ is shown in figure 2. Since $f(s \to 1) \to \infty$, this result applies if the second term is smaller than the first one. A last note is that at $s = 2$ the particle "falls" to the origin and the bound states disappear [1].

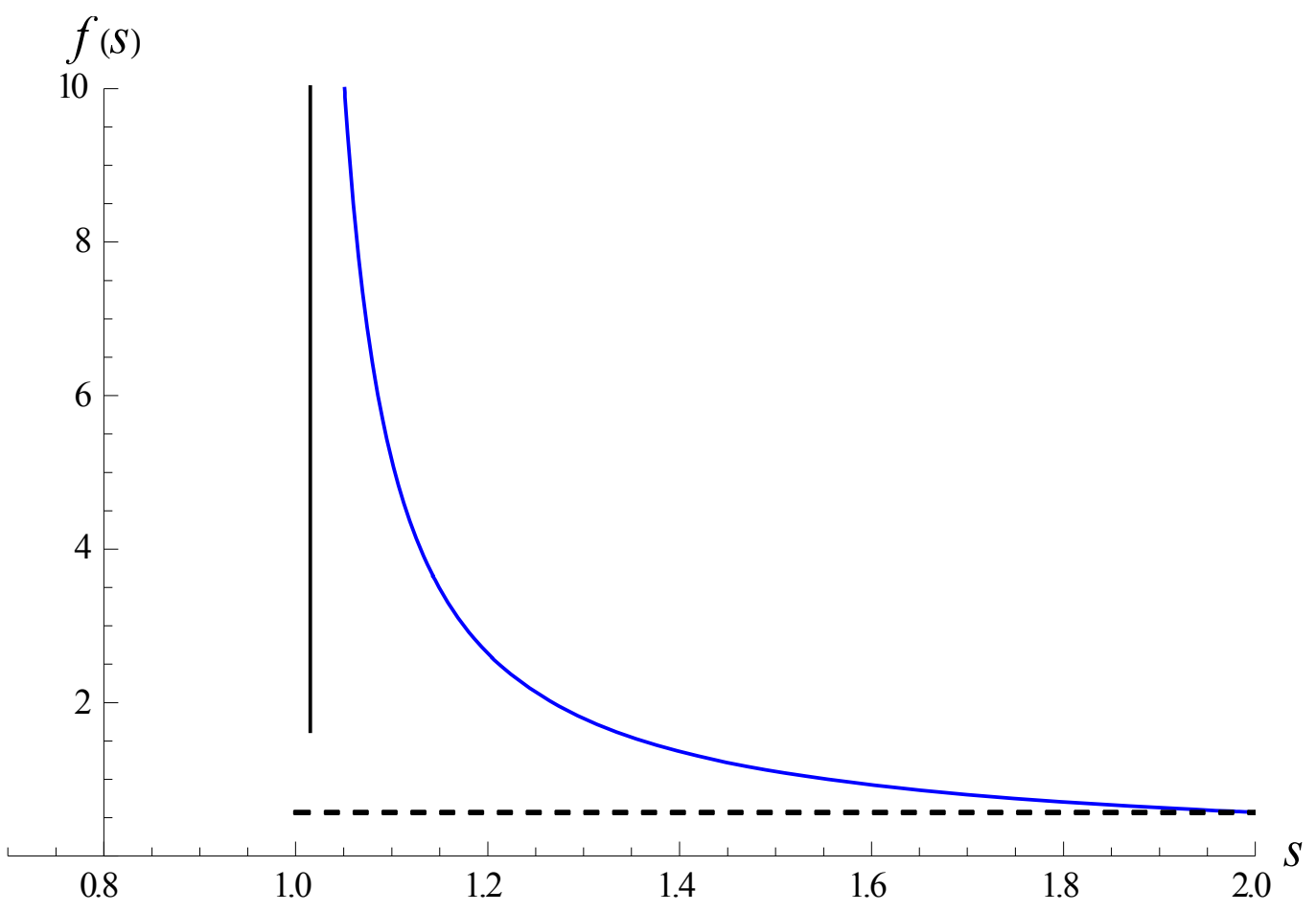

Fig. 2. The function $f(s)$, equation (19). The lower limit value $f(2) = -1 + \pi/2$ is shown by the dashed line.

### 3. Coulomb potential

Next we discuss the one dimensional Coulomb potential corresponding to $s = 1$. In this case (as well as for $s < 1$) the integral (19) diverges and the problem should be treated separately. The Coulomb energies are well-known: $E_n = -1/(2n^2)$, $n = 1, 2, 3...$ (in atomic units) [1]. They are the same as the WKB values. The WKB wave function of the unperturbed $n$-th state in the classically accessible region is

$$\phi_n(x) = \frac{A}{\sqrt{p_0(x)}} \cos\left( \int_x^{x_0} p_0(x')dx' - \frac{\pi}{4} \right), \tag{21}$$

where

$$p_0(x) = \sqrt{2\left( E_n + \frac{1}{x} \right)}. \tag{22}$$

The normalization factor according to (11) is

$$A^2 = \frac{2}{\pi n^3}. \tag{23}$$

Thus, the probability of the tunnel ionization per unit time from the $n$-th state in the one-dimensional Coulomb potential is

$$w = w_0 \exp\left( -2\int_{x_L}^{x_R} \sqrt{2\left(-E_n - 1/x - Fx\right)}\, dx \right), \quad w_0 = \frac{1}{2\pi n^3}, \tag{24}$$

where $E_n < 0$. With the transformation

$$x = \frac{z}{|E_n|}, \qquad \varepsilon = 4n^4 F > 0 \tag{25}$$

we have

$$I_s = -\frac{2}{|E_n|} \int_{z_1}^{z_2} \sqrt{2|E_n|\left(1 - \frac{1}{z} - \varepsilon z\right)}\, dz, \tag{26}$$

where $z_1$ and $z_2$ are the roots of the quadratic equation

$$z - 1 - \varepsilon z^2 = 0. \tag{27}$$

We expand them into power series:

$$z_1 = \frac{1 - \sqrt{1 - 4\varepsilon}}{2\varepsilon} = 1 + \varepsilon + O\left(\varepsilon^2\right), \tag{28}$$

$$z_2 = \frac{1 + \sqrt{1 - 4\varepsilon}}{2\varepsilon} = \frac{1}{\varepsilon} - 1 - \varepsilon + O\left(\varepsilon^2\right). \tag{29}$$

With the roots $z_1$ and $z_2$ the integral (26) is rewritten as

$$I_s = -\frac{2^{3/2}}{|E_n|^{1/2}} \int_{z_1}^{z_2} \sqrt{-\frac{\varepsilon}{z}\left(z - z_1\right)\left(z - z_2\right)}\, dz. \tag{30}$$

By the linear transformation $z = z_1 + \left(z_2 - z_1\right)t$ the limits of the integration are changed to 0 and 1:

$$I_s = -\frac{2^{3/2}}{|E_n|^{1/2}} \frac{\left(z_2 - z_1\right)^2}{\sqrt{z_1}} \varepsilon^{1/2} \int_0^1 \sqrt{\frac{t(1-t)}{1 - \frac{z_1 - z_2}{z_1} t}}\, dt. \tag{31}$$

The integral here is expressed in terms of the Gauss hypergeometric function:

$$I_s = -\frac{2^{3/2}}{|E_n|^{1/2}} \frac{\left(z_2 - z_1\right)^2}{\sqrt{z_1}} \varepsilon^{1/2} \frac{\pi}{8} {}_2F_1\left(\frac{1}{2}, \frac{3}{2}; 3; \frac{z_1 - z_2}{z_1}\right). \tag{32}$$

This is an exact result.

The formula (32) is invariant with respect to the transposition $z_1 \leftrightarrow z_2$. As a result of such a change the argument of the hypergeometric function belongs to the segment $[0,1]$ and is close to unity:

$$I_s = -\frac{2^{3/2}}{|E_n|^{1/2}} \frac{(z_2 - z_1)^2}{\sqrt{z_2}} \varepsilon^{1/2} \frac{\pi}{8} {}_2F_1\left(\frac{1}{2}, \frac{3}{2}; 3; 1 - \frac{z_1}{z_2}\right), \tag{33}$$

$$1 - \frac{z_1}{z_2} = 1 - \varepsilon + O(\varepsilon^2). \tag{34}$$

The expansion for small $\varepsilon$ then gives

$$I_s = -4n\left(\frac{2}{3\varepsilon} + \frac{\ln \varepsilon}{2} - \frac{\ln 16 + 1}{2} + O(\varepsilon)\right). \tag{35}$$

One can expand up to any order.

For the probability of tunnel ionization per unit time we get

$$w = \frac{1}{2\pi n^3}\left(\frac{4}{n^4 F}\right)^{2n} e^{-\frac{2}{3n^3 F} + 2n}. \tag{36}$$

When passing from the constant field to a low-frequency field the tunnel exponent should be averaged over the field period. Since

$$\begin{aligned} &\frac{\omega}{\pi} \int_{-\pi/2\omega}^{\pi/2\omega} \exp\left(-\frac{2(2|E_n|)^{3/2}}{3F\,|\cos \omega t\,|}\right) dt \approx \frac{\omega}{\pi} \int_{-\infty}^{\infty} \exp\left(-\frac{2(2|E_n|)^{3/2}}{3F}\left(1 + \frac{\omega^2 t^2}{2}\right)\right) dt = \\ &= \sqrt{\frac{3F}{\pi(2|E_n|)^{3/2}}} \exp\left(-\frac{2(2|E_n|)^{3/2}}{3F}\right), \end{aligned} \tag{37}$$

the probabilities obtained for the constant field should be multiplied by the factor [5]

$$\sqrt{\frac{3F}{\pi(2|E_n|)^{3/2}}}. \tag{38}$$

## 4. Inverse square root potential

Now we turn to the potential $V(x) = -1/\sqrt{x}$ considered in [6]. In the classically accessible region $1 \ll x < x_0$, where $x_0 = |E_n|^{-2}$, the unperturbed WKB wave function is that given by equation (10). The WKB energies of bound states are $E_n = -0.5(n - 1/6)^{-2/3}$ [3,4]. According to equation (11), in this case we have the normalization coefficient

$$A^2 = \frac{8\sqrt{2}}{3\pi} |E_n|^{5/2}. \tag{39}$$

The momentum when the particle moves below the barrier along the $x$ axis is

$$|p(x)| = \sqrt{2\left(|E_n| - \frac{1}{\sqrt{x}} - Fx\right)} \tag{40}$$

and the WKB wave function is the analytic continuation of the WKB function (10) into the region where the electric field becomes significant:

$$\phi(x) \to \frac{A}{2\sqrt{|p(x)|}} \exp\left(-\int_{x_0}^{x} |p(x')|\, dx'\right), \quad x_0 < x < x_1 = \frac{|E_n|}{F}. \tag{41}$$

The probability of the tunnel ionization per unit time is

$$w = \frac{A^2}{4} \exp\left(-2\int_{x_L}^{x_R} |p(x)|\, dx\right). \tag{42}$$

Let $x_L$ and $x_R$ be the left-hand and right-hand turning points. With the change $x = z^2 / |E_n|^2$ the integral

$$I_{1/2} = -2\int_{x_L}^{x_R} \sqrt{2\left(|E_n| - \frac{1}{\sqrt{x}} - Fx\right)} dx \tag{43}$$

is transformed to the form

$$I_{1/2} = -2\frac{2^{3/2}}{|E_n|^{3/2}} \int_{z_L}^{z_R} \sqrt{z\left(z - 1 - \varepsilon z^3\right)} dz\,, \tag{44}$$

where $\varepsilon = F / |E_n|^3$ is a small positive parameter. Let $z_{1,2,3}$ be the zeros of the cubic polynomial under the root in equation (44), so that

$$z - 1 - \varepsilon z^3 = -\varepsilon(z - z_1)(z - z_2)(z - z_3)\,. \tag{45}$$

Using Cardano's formula, we readily get the following expansion for $z_{1,2,3}$ for small $\varepsilon \ll 1$:

$$z_{1,2,3} = \left(-\frac{1}{2} - \frac{1}{\sqrt{\varepsilon}} + \frac{3\sqrt{\varepsilon}}{8} - \frac{\varepsilon}{2} + \ldots,\ 1 + \varepsilon + \ldots,\ -\frac{1}{2} + \frac{1}{\sqrt{\varepsilon}} - \frac{3\sqrt{\varepsilon}}{8} - \frac{\varepsilon}{2} + \ldots\right). \tag{46}$$

These roots are arranged in ascending order, and it holds $z_1 < 1 < z_2 < z_3$. The left-hand turning point is $z_L = z_2$, and the right-hand one is $z_R = z_3$. So, the integral (44) is rewritten as

$$I_{1/2} = -\frac{4\sqrt{2}\varepsilon}{\sqrt{F}} \int_{z_2}^{z_3} \sqrt{-z(z - z_1)(z - z_2)(z - z_3)}\, dz\,. \tag{47}$$

This integral is expressed in terms of the Appell generalized hypergeometric function of two variables of the first kind [7]. This is achieved by the linear transformation

$$z = z_2 + (z_3 - z_2)t\,, \tag{48}$$

which reduces $I_{1/2}$ into an integral with integration limits 0 and 1:

$$I_{1/2} = -\frac{4\sqrt{2}\varepsilon}{\sqrt{F}}\left(z_3 - z_2\right)\sqrt{z_2(z_2 - z_1)}\int_0^1 \sqrt{t(1-t)(1-y_1 t)(1-y_2 t)}\,dt\,, \qquad (49)$$

where

$$y_{1,2} = \left(1 - \frac{z_3}{z_2}, \frac{z_3 - z_2}{z_1 - z_2}\right). \qquad (50)$$

The Appell generalized hypergeometric function of two variables of the first kind $F_1(y_1, y_2)$ has the following integral representation [7]:

$$F_1(a; b_1, b_2; c; y_1, y_2) = \frac{\Gamma(c)}{\Gamma(a)\Gamma(c-a)}\int_0^1 t^{a-1}\left(1-t\right)^{c-a-1}\left(1-y_1 t\right)^{-b_1}\left(1-y_2 t\right)^{-b_2} dt\,, \qquad (51)$$

which is correct if $\mathrm{Re}(a) > 0$ and $\mathrm{Re}(c-a) > 0$. Comparing with equation (49), we get $(a, b_1, b_2, c) = (3/2, -1/2, -1/2, 3)$ and, thus,

$$I_{1/2} = -\frac{\pi\sqrt{2}V_0\varepsilon}{2\sqrt{F}}\left(z_2 - z_3\right)^2\sqrt{z_2(z_2 - z_1)}\,F_1\left(\frac{3}{2}; -\frac{1}{2}, -\frac{1}{2}; 3; 1 - \frac{z_3}{z_2}, \frac{z_3 - z_2}{z_1 - z_2}\right). \qquad (52)$$

This is the exact value of the exponent in the WKB approximation for the tunnel ionization rate (42) for the inverse square root potential $-V_0/\sqrt{x}$:

$$w = \frac{2\sqrt{2}}{3\pi}\left|E_n\right|^{5/2}\exp(I_{1/2}). \qquad (53)$$

We note that $I_{1/2}$ is a function of one variable $\varepsilon$. Together with Cardano's formula for the zeros of a cubic polynomial, it permits calculation of the semiclassical probability of the tunnel ionization for any given accuracy. Consider the asymptotic expansion of $I_{1/2}(\varepsilon)$ for small $\varepsilon$. Using the identities obeyed by the Appell functions [7], we note that $I_{1/2}$ is invariant with respect to the transposition $z_2 \leftrightarrow z_3$. This transposition makes the first argument of the Appell function be close to the unity and the second one – to 1/2. Thus,

$$I_{1/2} = -\frac{\pi\sqrt{2}V_0\varepsilon}{2\sqrt{F}}\left(z_3 - z_2\right)^2\sqrt{z_3(z_3 - z_1)}\,F_1\left(\frac{3}{2}; -\frac{1}{2}, -\frac{1}{2}; 3; 1 - \frac{z_2}{z_3}, \frac{z_2 - z_3}{z_1 - z_3}\right) \qquad (54)$$

and $$y_1 = 1 - \frac{z_2}{z_3} \approx 1 - \sqrt{\varepsilon} - \frac{\varepsilon}{2} + O\left(\varepsilon^{3/2}\right), \quad y_2 = \frac{z_2 - z_3}{z_1 - z_3} \approx \frac{1}{2} - \frac{3\sqrt{\varepsilon}}{4} + O\left(\varepsilon^{3/2}\right). \qquad (55)$$

In the vicinity of the point $y_1 = 1$ the Appell function admits an expansion in terms of the hypergeometric functions. The first two terms read

$$F_1 = {}_2F_1(a,b_1;c;1)\,{}_2F_1(a,b_2;c-b_1;y_2) + \frac{ab_1}{c}\,{}_2F_1(a+1,b_1+1;c+1;1)\,{}_2F_1(a+1,b_2;c-b_1;y_2)(y_1-1) + O\left((y_1-1)^2\right), \tag{56}$$

which gives

$$F_1 = \frac{32}{15\pi}\,{}_2F_1\left(-\frac{1}{2},\frac{3}{2};\frac{7}{2};y_2\right) + \left(\sqrt{1-y_2}\left(\frac{2}{y_2}+\frac{3}{y_2^2}-8\right) - \frac{3\sin^{-1}\left(\sqrt{y_2}\right)}{y_2^{5/2}}\right)\frac{(y_1-1)}{6\pi} + O\left((y_1-1)^2\right). \tag{57}$$

Expanding this function in terms of $\varepsilon$ and substituting into equation (54), we finally get

$$I_{1/2} \approx -\frac{4V_0\varepsilon}{3\sqrt{F}}(z_3-z_2)^2\sqrt{z_3(z_3-z_1)}\left(1+\frac{(13-3\pi)\sqrt{\varepsilon}}{4}+\frac{(283-84\pi)\varepsilon}{32}\right). \tag{58}$$

Using Cardano's formula, the pre-factor before the brackets is also expanded into series in terms of half-integer powers of $\varepsilon$:

$$(z_3-z_2)^2\sqrt{z_3(z_3-z_1)} = \frac{\sqrt{2}}{\varepsilon^{3/2}} - \frac{13}{2\sqrt{2}\varepsilon} + \frac{59}{16\sqrt{2}\sqrt{\varepsilon}} - \frac{165}{64\sqrt{2}} + O\left(\varepsilon^{1/2}\right). \tag{59}$$

Eventually, we get a rather simple expansion:

$$I_{1/2} = -\frac{4\sqrt{2}V_0\varepsilon}{\sqrt{F}}\left(\frac{1}{3\varepsilon^{3/2}} - \frac{\pi}{4\varepsilon} + \frac{2-3\pi}{48\sqrt{\varepsilon}} + O(1)\right). \tag{60}$$

Substituting this into equation (53), we get the result

$$w = \frac{2\sqrt{2}|E_n|^{5/2}}{3\pi}\exp\left(-\frac{2\left(2|E_n|\right)^{3/2}}{3F} + \frac{\pi\sqrt{2}}{\sqrt{F}} - \frac{\sqrt{2}(2-3\pi)}{12|E_n|^{3/2}}\right). \tag{61}$$

## 5. Logarithmic potential

As it is seen from the previous analysis, the role of the right-hand turning point in the tunnel ionization increases with the decrease of the degree $s$. We conclude with consideration of the logarithmic potential $V(x) = V_0\ln(x/a)$, $V_0 > 0$, for which this role is most pronounced since this potential corresponds to the limit $s \to 0$. The WKB energies of the bound states in this potential are [3,4]:

$$E_n = V_0\ln\left(\frac{(n-1/4)}{a}\sqrt{\frac{2\pi}{V_0}}\right), \quad n = 1,2,3... \tag{62}$$

The WKB wave-function in the classically accessible region (but not too close to the origin) has the form

$$\phi_n(x) = \frac{A}{\sqrt{p_0(x)}} \cos\left( \int_x^{x_L} p_0(x') dx' - \frac{\pi}{4} \right), \quad A^2 \int_0^{x_L} \frac{dx}{2 p_0(x)} = 1, \tag{63}$$

$$p_0(x) = \sqrt{2\left( E_n - V_0 \ln \frac{x}{a} \right)}, \tag{64}$$

where $x_L$ is the left-hand turning point under the barrier:

$$x_L \approx (n - 1/4) \sqrt{\frac{2\pi}{V_0}}. \tag{65}$$

Substituting equations (64) and (65) into equation (63) and integrating, we get the normalization factor $A$:

$$A^2 \approx \frac{2V_0}{\pi(n - 1/4)}. \tag{66}$$

The probability of the tunnel ionization per unit time is given as

$$w = \frac{V_0}{\pi(n - 1/4)} \exp\left( -2 \int_{x_L}^{x_R} |p(x)| dx \right), \tag{67}$$

$$|p(x)| = \sqrt{2\left( -E_n + V_0 \ln \frac{x}{a} - Fx \right)}. \tag{68}$$

For the right-hand turning point $x_R$ we have

$$x_R \approx \frac{V_0}{F} \ln \frac{1}{\varepsilon}, \quad \varepsilon = \frac{(n - 1/4) F \sqrt{2\pi}}{V_0^{3/2}} \ll 1. \tag{69}$$

First, we estimate the main contribution to the probability of the tunnel ionization. This is the contribution from the vicinity of the right-hand turning point:

$$I \approx \int_0^{x_R} \sqrt{F(x_R - x)} dx = \frac{2V_0^{3/2}}{3F} \left( \ln \frac{1}{\varepsilon} \right)^{3/2}. \tag{70}$$

For the leading term in the tunnel ionization rate we then have

$$w = \frac{V_0}{\pi(n - 1/4)} \exp\left( -\frac{4\sqrt{2} \cdot V_0^{3/2}}{3F} \left( \ln \frac{V_0^{3/2}}{(n - 1/4) F \sqrt{2\pi}} \right)^{3/2} \right). \tag{71}$$

According to (38), the effect of a low-frequency field in this case results in an additional pre-exponential factor

$$\left(\ln \frac{V_0^{3/2}}{(n-1/4)F\sqrt{2\pi}}\right)^{-3/4} \sqrt{\frac{3F}{\pi(2V_0)^{3/2}}}. \tag{72}$$

This obtained result can be improved in the following way. The transformation

$$x = \frac{V_0 \varepsilon}{F} e^z, \quad \varepsilon = \frac{aF}{V_0} e^{E_n/V_0} \tag{73}$$

reduces the integral in equation (67) to the form

$$I = -\frac{(2V_0)^{3/2}\varepsilon}{F} \int_{z_L}^{z_R} e^z \sqrt{z - \varepsilon e^z}\, dz. \tag{74}$$

The turning points are defined by the equation $z = \varepsilon e^z$. The result reads

$$z_L = -W(-\varepsilon), \quad z_R = -W_{-1}(-\varepsilon), \tag{75}$$

where $W$ and $W_{-1}$ are the upper and lower branches of the Lambert function [8,9]. For small $\varepsilon \ll 1$ the following expansions are valid:

$$z_L = \varepsilon + \varepsilon^2 + \ldots, \quad z_R = \ln\frac{1}{\varepsilon} + \ln\left(\ln\frac{1}{\varepsilon}\right) + \ldots. \tag{76}$$

The exponential factor in the integrand in equation (74) indicates that the main contribution to the integral is indeed made by the right-hand turning point. So, we expand the integrand in the vicinity of that point:

$$e^z\sqrt{z - \varepsilon e^z} = e^z\sqrt{z_R - \varepsilon e^z} + \frac{e^z(z - z_R)}{2\sqrt{z_R - \varepsilon e^z}} + O\left((z - z_R)^2\right) + \ldots, \tag{77}$$

This is a rather accurate expansion well approximating the integrand function on the whole segment $[z_L, z_R]$. If the second terms in the expansions (76), (77) are not taken into account and the contribution from the left-hand turning point is neglected (that is the lower integration limit is set to 0), we get the result (70). Consideration of the second terms gives

$$I \approx -\frac{(2V_0 z_R)^{3/2}}{3F}\left(1 + \frac{6}{z_R}\left(1 - \tanh^{-1}\left(\sqrt{1 - \frac{z_L}{z_R}}\right)\right)\right). \tag{78}$$

The accuracy of this result in comparison with equation (70) is demonstrated in figure 3. It is understood that to get the correct limit of the integral at $\varepsilon \to 0$ more accurate determination of the turning points and better approximation of the integrand are necessary.

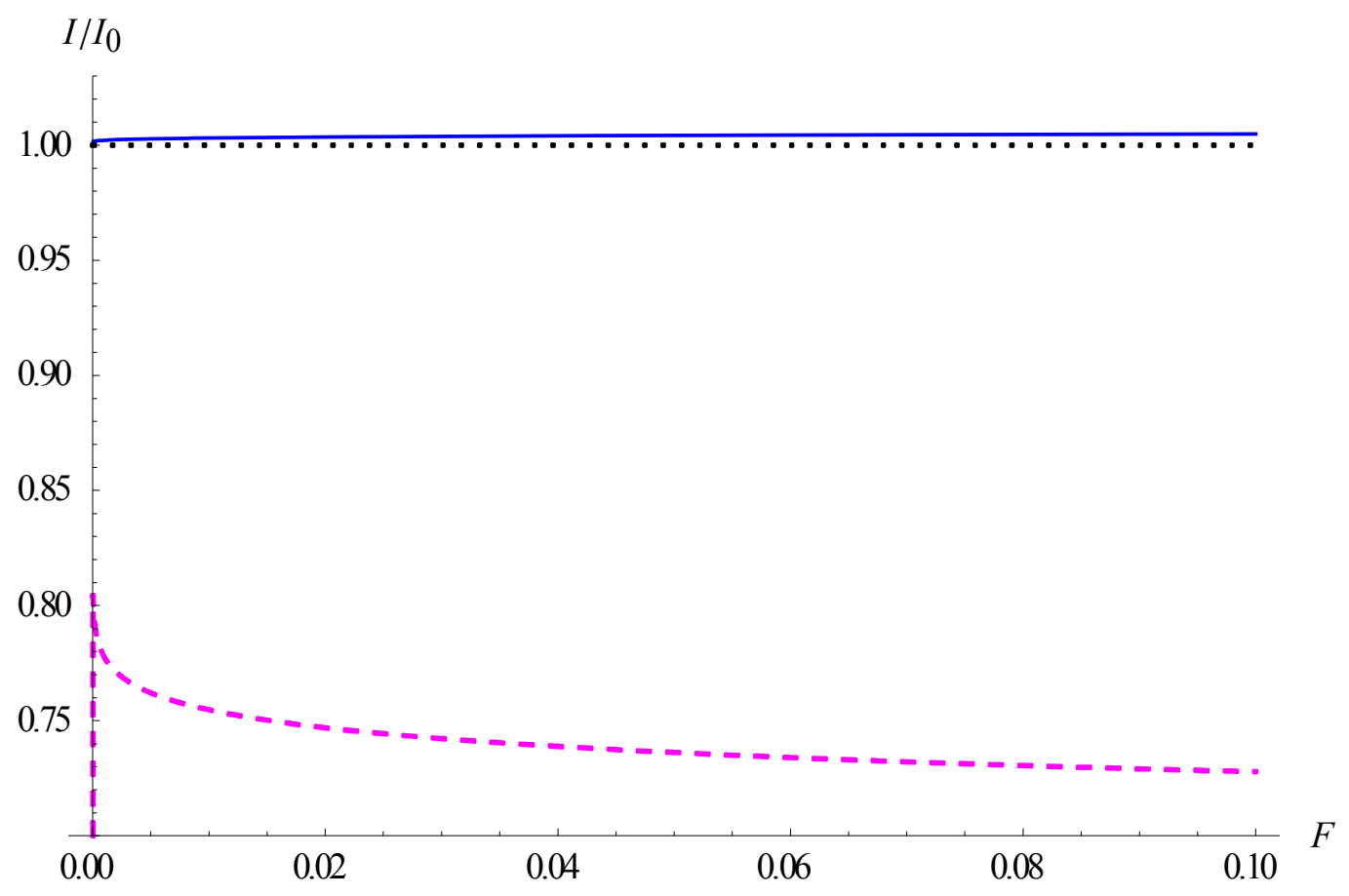

Fig. 3. Integrals (78) (upper solid line) and (70) (lower dashed line), normalized to the exact numerical value $I_0$ of the integral (dotted line)

## 6. Discussion

The approach we used to calculate the tunnel ionization rate in a constant or low-frequency electric field can also be applied to treat other one-, two- or three-dimensional potentials, for instance, non-singular power-law potentials of a positive exponent: $V(x)=V_0x^s$ with $1>s>0$, $V_0>0$, the Rosen-Morse potential $V(x)=-V_0/\cosh^2(x/a)$, $V_0>0$ [10], etc. Many advanced potentials, the exact solutions for which are written in terms of the Heun functions [11], can also be treated in the same manner. In particular, of a certain interest is the short-range singular Lambert-W potential, which behaves like the inverse square root potential in the vicinity of the origin and decreases exponentially at infinity [12]. The Lennard-Jones composite power-law potential $V(x)=A/x^{12}-B/x^6$ [13] is another example of a two-scale potential that suggests distinct behavior of the turning points.

In the cases, when the barrier penetrability is expressed through known special functions (including the advanced functions of the Heun class [14]), the problem becomes easier since the known asymptotic expansions of a given order for these functions can be applied as it was done here for the Gauss hypergeometric and the Appell generalized hypergeometric functions. Otherwise one may confine himself to the leading term in the penetrability as it was done for the logarithmic potential. Although we have considered only the WKB states, it is known that the accuracy of the latter wave-functions is rather high for both large and small quantum numbers and even for the ground state if the Maslov index is added to the principal quantum number [4].

This research was supported by the Ministry of Education and Science of the Russian Federation (project № 3.873.2017/К), the State Committee of Science of the Republic of Armenia (project 15T-1C323), the Armenian National Science and Education Fund (ANSEF grant PS-4558), as well as by the project “Leading research universities of Russia” (grant FTI_24_2016 of the Tomsk Polytechnic University).